\documentclass[a4paper]{jpconf}

\usepackage{graphicx}
\usepackage[english]{babel}
\usepackage{color}
\usepackage{amsmath,amssymb,amsfonts}

\newcommand{\ener}[1]{ \varepsilon_{#1} }
\newenvironment{eq}{\begin{eqnarray}}{\end{eqnarray}}
\newenvironment{fig}{\begin{figure}[tp]\begin{minipage}{\textwidth}\centering}{\end{minipage}\end{figure}}

\graphicspath{{./fig/}}

\begin{document}
\title{Dark Matter detection via lepton cosmic rays}
\author{Roberto A. Lineros}
\address{INFN, sezione di Torino, I-10122 Torino, Italy and
Dipartimento di fisica teorica, Universit\`a di Torino, I-10122 Torino,
Italy.}
\ead{lineros@to.infn.it}
\begin{abstract}
Recent observations of lepton cosmic rays, coming from the PAMELA and FERMI experiments, have pushed our understanding of the interstellar medium and cosmic rays sources to unprecedented levels. The imprint of dark matter on lepton cosmic rays is the most exciting explanation of both PAMELA's positron excess and FERMI's total flux of electrons. 
Alternatively, supernovae are astrophysical objects with the same potential to explain these observations. In this work, we present an updated study of the astrophysical sources of lepton cosmic rays and the possible trace of a dark matter signal on the positron excess and total flux of electrons.
\end{abstract}

\begin{flushleft}
  {\scriptsize Proceeding for the 16th International Symposium on Particles, Strings and Cosmology. PASCOS 2010, Valencia, Spain. \\}
  {\scriptsize Preprint DFTT 16/2010 }
\end{flushleft}

\section{Introduction}
Last decades, different cosmic ray experiments like HEAT~\cite{1997ApJ...482L.191B}  and AMS01~\cite{2000PhLB..484...10A} have presented very interesting cosmic rays results, specially for cosmic rays of \emph{electrons}\footnote{in this work, we refer to electrons and positrons as \emph{electrons}}, in which the theoretical prediction of the positron fraction (number of positron divided by the number of \emph{electrons} per unit of energy) based on standard sources mismatched the experimental results, and observing an increment in the fraction for energies larger than 10~GeV.
Recently, newer and refined experiments like PAMELA~\cite{Adriani:2008zr} and FERMI~\cite{2010arXiv1008.3999F, 2010arXiv1008.5119T} present more accurate measurements of the positron fraction and the \emph{electrons} flux, confirming the positron excess and other kinds of anomalies present in the observations which are not been completely explained by theoretical predictions.\\

In this work, we analyze the potential of dark matter and astrophysical sources to explain these anomalies. 
Moreover, we consider the effect of uncertainties in the propagation model and in the sources to verify the robustness of a possible discovery.\\
 

\section{Propagation of cosmic rays}

Cosmic rays propagation is rather different of propagation of photons and neutrinos, which is a straight line among the source and the observer.
Cosmic rays have to deal with inhomogeneous magnetic fields, radiation fields from stellar activity and from the microwave background, and with the interstellar gas.  
The continuous interaction with the environment makes them to loose (or gain) energy and to propagate following a diffusion pattern.\\

Precise description of possibles cosmic ray interaction with the environment is far to be reached due to the uncertainties related to  environmental quantities in the whole Galaxy. 
For instance, the galactic center is a very dynamic region that makes hard to model it and to provide under-control estimations.
In opposition, regions outside the Galactic disk -- which are very important for cosmic ray propagation -- have few astrophysical objects, and then big uncertainties in the estimation of magnetic fields, radiation field, etc.\\

The model of propagation of cosmic rays is based on a continuity equation for the number density of cosmic rays, known as transport equation, which contains most of the processes between cosmic rays and the environment.\\
In our case, the transport equation for cosmic rays of electrons and positrons is:
\begin{eq}
  \label{eq:std_te}
  \frac{\partial \psi}{\partial t}  - \nabla \big( D(\ener{}) \nabla \psi \big) - \frac{\partial}{\partial \ener{}} \big(b(\ener{})
\psi \big) = q(\vec{x},\ener{}) \, ,
\end{eq}
where $\psi$ is the number density of electrons or positrons per unit of energy, $D(\ener{})$ is the diffusion term, $b(\ener{})$ the energy loss term, and $q(\vec{x},\ener{})$ is the source term.\\

The diffusion term is considered homogeneous in space with a power laws dependence in energy:
\begin{eq}
	D(\ener{}) = K_0 \, \left(\frac{\ener{}}{\ener{0}}\right)^{\delta} \, ,
\end{eq}
where $K_0$ and $\delta$ are phenomenological parameters inspired by magnetohydrodynamics models of the interstellar medium. 
The normalization energy scale $\ener{0}$ here is fixed at the value of 1~GeV.\\

The energy loss term depends on the interaction between the galactic environment and the type of cosmic rays.
The process that dominates the energy evolution of electrons and positrons at GeV--TeV scale is the inverse Compton scattering with the Interstellar Radiation Field and the synchrotron emission related to the Galactic magnetic field.\\
For simplicity, the inverse Compton scattering can be used in the Thomson regime which is valid for \emph{electrons} with energy lower than the equivalent photon energy (details in \cite{2010arXiv1002.1910D,1970RvMP...42..237B}). 
In this case, the energy loss term corresponds to:
\begin{eq}
b(\ener{}) = \frac{\ener{0}}{\tau_E} \left(\frac{\ener{}}{\ener{0}}\right)^2 \, , 
\end{eq}
where $\tau_E$ is the energy loss scale time which is calculated from the total photon energy density.\\ 
However, this formula is not accurate for high energy \emph{electrons} due to the scattering with the photon bath becomes less frequent.
This regime is known as Klein-Nishina regime~\cite{springerlink:10.1007/BF01366453} and becomes non-negligible for the interaction of GeV-TeV \emph{electrons} with the ultraviolet, infrared and microwave component of the interstellar radiation field~\cite{1970RvMP...42..237B, 2010arXiv1002.1910D}.\\


%
For generic functions of the diffusion and energy loss terms,  solutions of the transport equation (Equation \ref{eq:std_te}) are
fully analytical. 
Analytical solutions for continuous and burst injection cases (details in \cite{2008arXiv0812.4272L,2010arXiv1002.1910D}) are fully
of physical meaning.
We remark that the analytical approach allows us to study the importance of different sources and to scan the propagation
space of parameters in rather short time compared to more sophisticated methods.\\

\subsection{Uncertainties}
Most of the physical processes are contained in the propagation model.
Nevertheless, these processes depend on environmental variables that are constrained by observations but not completely determined.\\

The principal propagation parameters are the diffusion parameters ($K_0$, $\delta$), and the geometry of the zone of propagation.
It is usually a cylinder of radius equal to the galactic one (20 kpc) and half-thickness $L_z$ which is a parameter of the model \cite{1980Ap&SS..68..295G}. 
The model also includes the escape of cosmic rays from the propagation zone by imposing that cosmic rays density equal to zero at the boundaries. 
As well, the escape of cosmic rays can be archived when a \mbox{non-homogeneous} diffusion term is considered.
Some models~\cite{2009arXiv0909.4548D} use a diffusion term that grows exponentially,
\begin{eq}
D_{\exp}(z,\ener{}) = D(\ener{}) \exp\left(\frac{|z|}{z_t}\right)  \, ,
\end{eq}
where $z_t$ is a vertical scale. 
This family of models provides a smooth transition among the diffusive and the straight--line propagation zones but loosing analyticity and needs to be solved numerically.\\ 

One method to constrain the propagation space of parameter is based on observation of the Boron/Carbon ratio (B/C)  \cite{2001ApJ...555..585M}. 
For the scope of our analysis, we assume that propagation space of parameters is common for all species of cosmic rays.\\

\section{Sources of cosmic rays electrons}

As important as the propagation, sources of \emph{electrons} present a challenge for models of production and injection.
In principle, each source is different from the others in the following aspects: spatial and temporal distribution, energy spectrum, and intensity.
Also, if we consider sources in the Galaxy, physical processes regarding to cosmic rays acceleration, nuclear physics, and physics beyond standard model would have an impact into the observed flux of \emph{electrons}. \\

At this point, we describe some of the principal sources of \emph{electrons} at galactic scale.
Our intention is to show how the observations by PAMELA and FERMI would be described as different combination of sources.

\subsection{Dark matter}
\begin{fig}
\centering
\resizebox{0.8\hsize}{!}{\includegraphics[angle=-90]{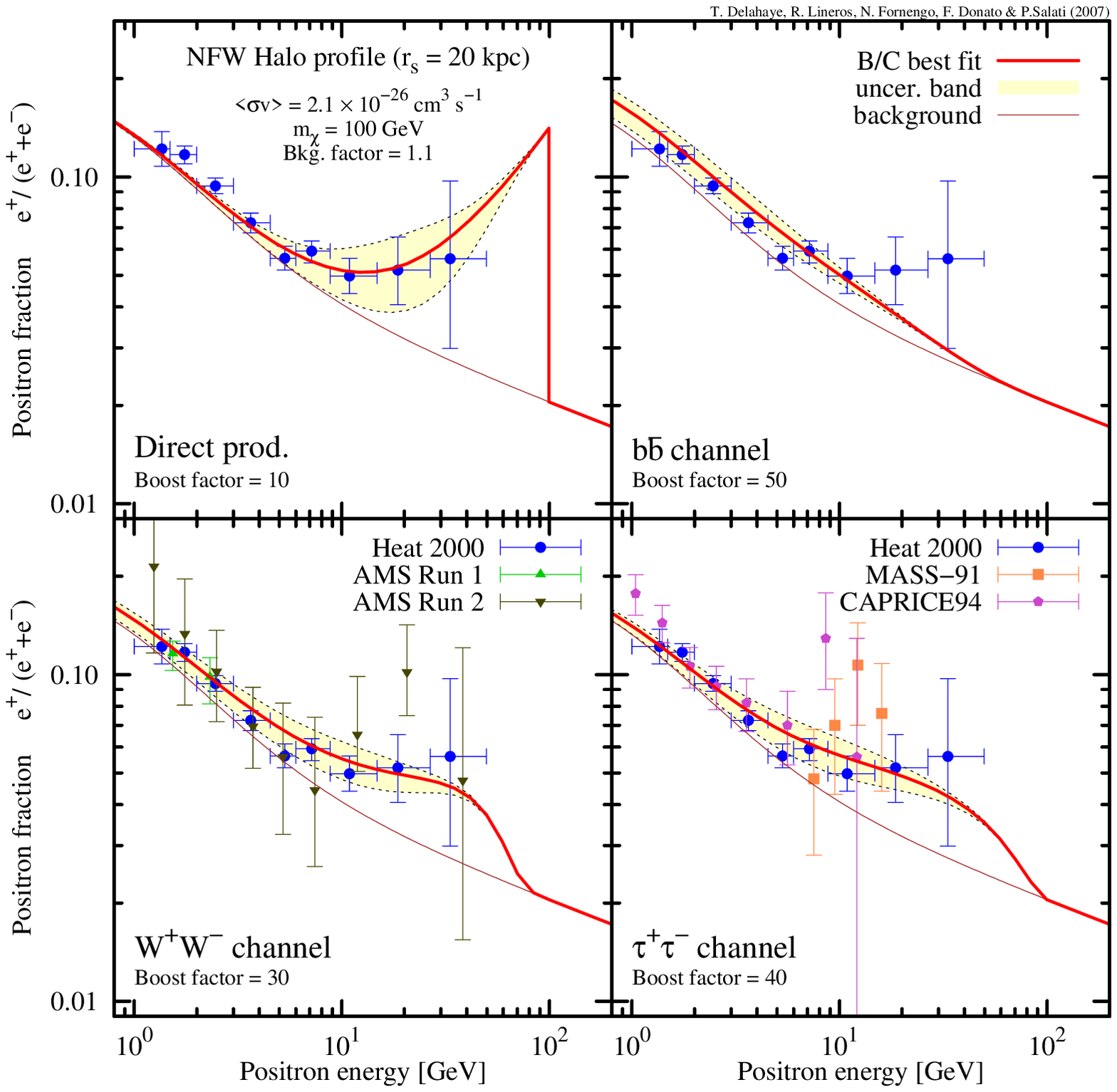}\includegraphics[angle=-90]{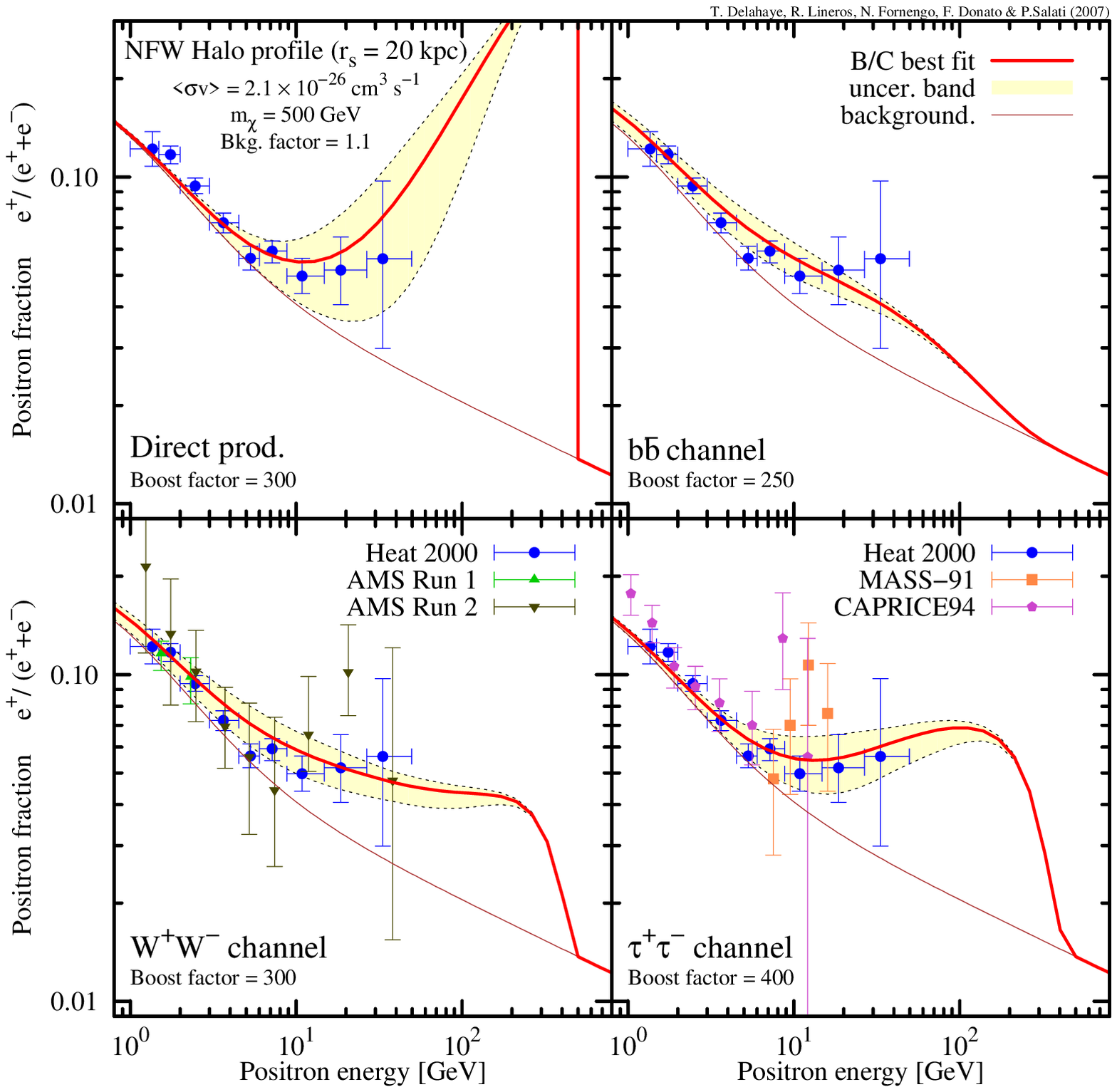}}
\caption{\label{fig1} 
Positron fraction versus energy for 4 different annihilation channels: direct production of positrons, $b\bar{b}$, $W^{+}W^{-}$, and $\tau^{+}\tau^{-}$.  
(Left-panel) case with dark matter mass of 100~GeV. (Right-panel) case with dark matter mass of 500~GeV.
Yellow bands in each plot corresponds to different propagation models compatible with B/C analysis.
Details in Ref. \cite{2008PhRvD..77f3527D}.}
\end{fig}
%
%

This is one of the most interesting and active topic.
The possibility that cosmic rays anomalies are due to the annihilation or decay of dark matter particles it is far to be the most excited solution to the puzzle.
The evidences obtained from cosmology and galaxy dynamics, among other,  point towards the presence of dark matter. However, its imprints  in cosmic rays, photons, neutrinos, etc. are not well defined, yet. \\

In the case of annihilation of dark matter, the source term of \emph{electrons} is:
\begin{eq}
  q_{\textnormal{DM}}(\vec{x},\ener{}) = \eta \langle \sigma v\rangle \, \left( \frac{\rho_{\textnormal{DM}}(\vec{x})}{m_{\chi}} \right)^2 \frac{dn}{d\ener{}} (\ener{}) \, ,
\end{eq}
where $\eta$ is a statistical coefficient which depends on dark matter nature, $\langle \sigma v\rangle$ is thermally averaged cross section which related with the dark matter relic abundance, $\rho_{\textnormal{DM}}$ is the dark matter distribution, $m_{\chi}$ is the dark matter mass, and $\frac{dn}{d\ener{}}$ is the multiplicity distribution of electrons or positrons produced in an annihilation event.\\

In principle, the signal from annihilation of dark matter depends on  the annihilation channel and the dark matter mass. 
In figure~\ref{fig1}, we presents the positron fraction calculated for different annihilation channels and masses~\cite{2008PhRvD..77f3527D}. 
We note that channels which produce mainly leptons are favorite to explain the signal. 
In addition, the  yellow bands represent  the propagation models compatible with the B/C analysis. 
It is remarkable that a possible dark matter signal is not vanished by the uncertainties in the propagation~\cite{2008PhRvD..77f3527D}.
Nevertheless,  uncertainty in the propagation should also affect the background of \emph{electrons},  which should be from astrophysical origin.\\

\subsection{Secondaries}
\begin{fig}
\centering
\resizebox{0.45\hsize}{!}{\includegraphics{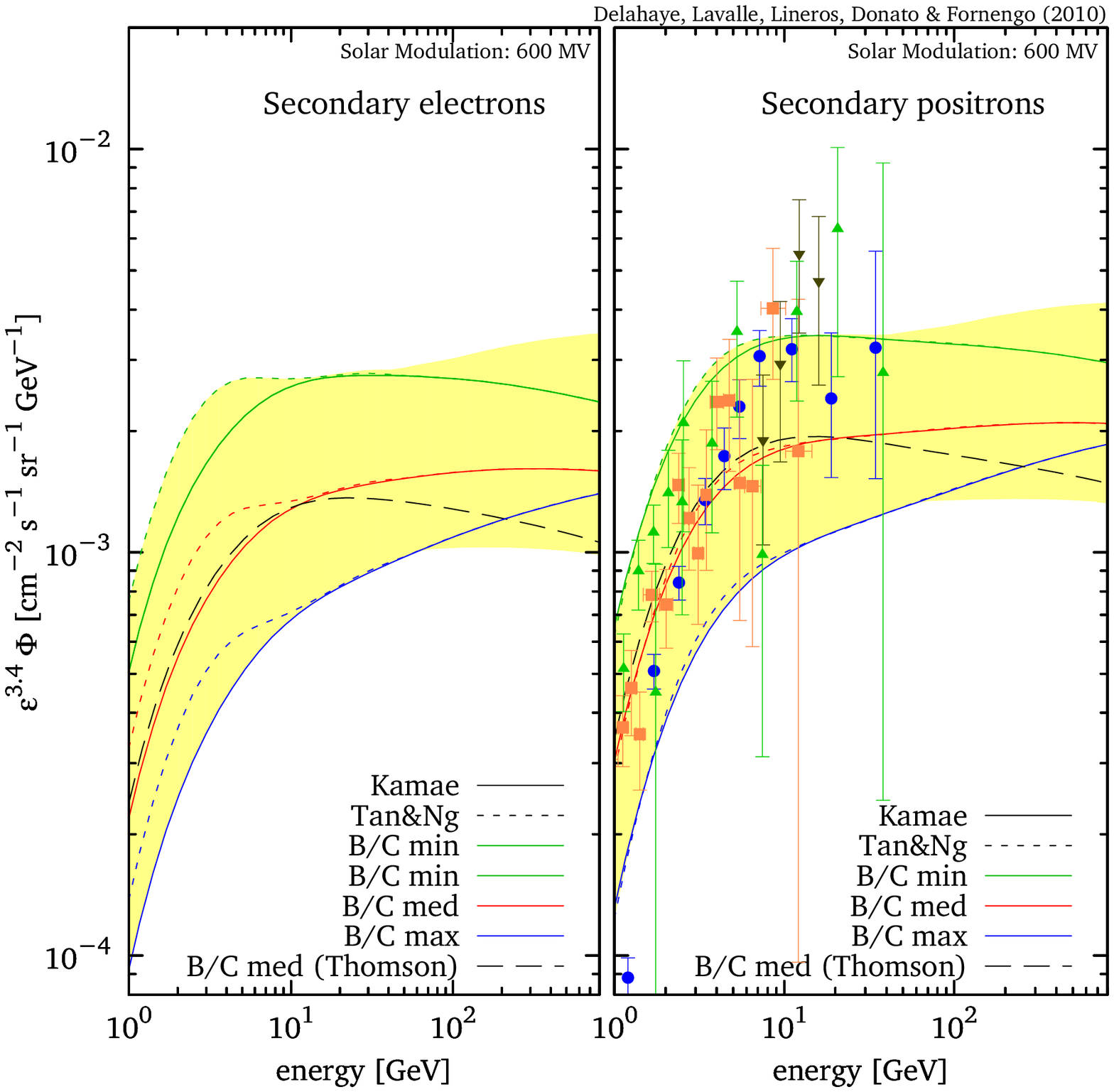}}
\caption{\label{fig2} 
Secondary electrons (left) and positrons (right) versus energy.
Yellow bands correspond to propagation models compatible with B/C analysis.
Secondary positrons encompass the available data which is expected at low energies.
Details in~\cite{2010arXiv1002.1910D,2009A&A...501..821D}.}
\end{fig}
This component is the result of the spallation of nuclear cosmic rays (protons and alpha particles) on the interstellar gas, which is mainly composed by hydrogen and helium.\\
In figure~\ref{fig2}, we present the calculation for both type of secondaries.
Electrons from secondary processes are not dominant in the electron flux instead of positrons that encompass the available low energy flux~\cite{2009A&A...501..821D}. 
As in the dark matter case, yellow bands correspond to the compatible B/C space of parameter.
We denote that the compatibility among observation and propagation space of parameter reaffirm the idea that propagation models are common for every specie of cosmic rays.\\

\subsection{Supernovae remnants and pulsars}

Supernovae are the astrophysical objects responsible of most of the galactic cosmic ray.
Moreover, those may give birth to two astrophysical objects with potential to produce electrons and positrons: Supernova remnants and pulsars.\\
Supernova remnants can expel a big fraction of the electron cosmic rays which is contained inside the former star, but very few positrons. 
Different mechanisms have been proposed to enhance the positron production~\cite{2009PhRvL.103e1104B}, however, it seems to be not enough to explain the both anomalies.
On the other hand, pulsars are able to produce in same amount positrons and electrons due to the interaction of pulsar's magnetic field with ambient photons. \\

In both cases, the injection spectra is expected to follow a power-law like function:
\begin{eq}
  q_{\textnormal{SN}} \propto Q_{0} \ener{}^{-\gamma} \exp\left(- \frac{\ener{}}{\ener{c}} \right) ,
\end{eq}
where $Q_{0}$ is the injection intensity of electrons/positrons, which is calculated from the averaged energy released by the supernova (or pulsar) and the efficiency to convert this energy into electron kinetic energy (details in~\cite{2010arXiv1002.1910D}), $\gamma$ is the power index, and $\ener{c}$ is a cut-off energy ($~$1--10~TeV) suggested by FERMI and HESS~\cite{2008PhRvL.101z1104A}.\\

Let us remark that supernovae are distributed \mbox{non-smoothly} in space and time.
Nevertheless, the diffusive behavior of propagated \emph{electrons} smooths any peak at low energy, because this range is dominated by the older and farther supernovae.
On the other hand, it is expected to be observed some features at high energies due to the contribution of younger and closer objects.\\
%

In figure~\ref{fig3}, we present some of the results of a detailed study of astrophysical sources of \emph{electrons} at the GeV-TeV scale~\cite{2010arXiv1002.1910D}.
We remark that under conservative assumption of injection profiles, the observations by FERMI and PAMELA can be encompassed with just the astrophysical component i.e. supernova remnants, pulsars and secondaries.
Also, the presence of inhomogeneities in the local source distribution naturally explains the features observed by FERMI at the TeV range. \\

\begin{fig}
\centering
\resizebox{0.8\hsize}{!}{\includegraphics{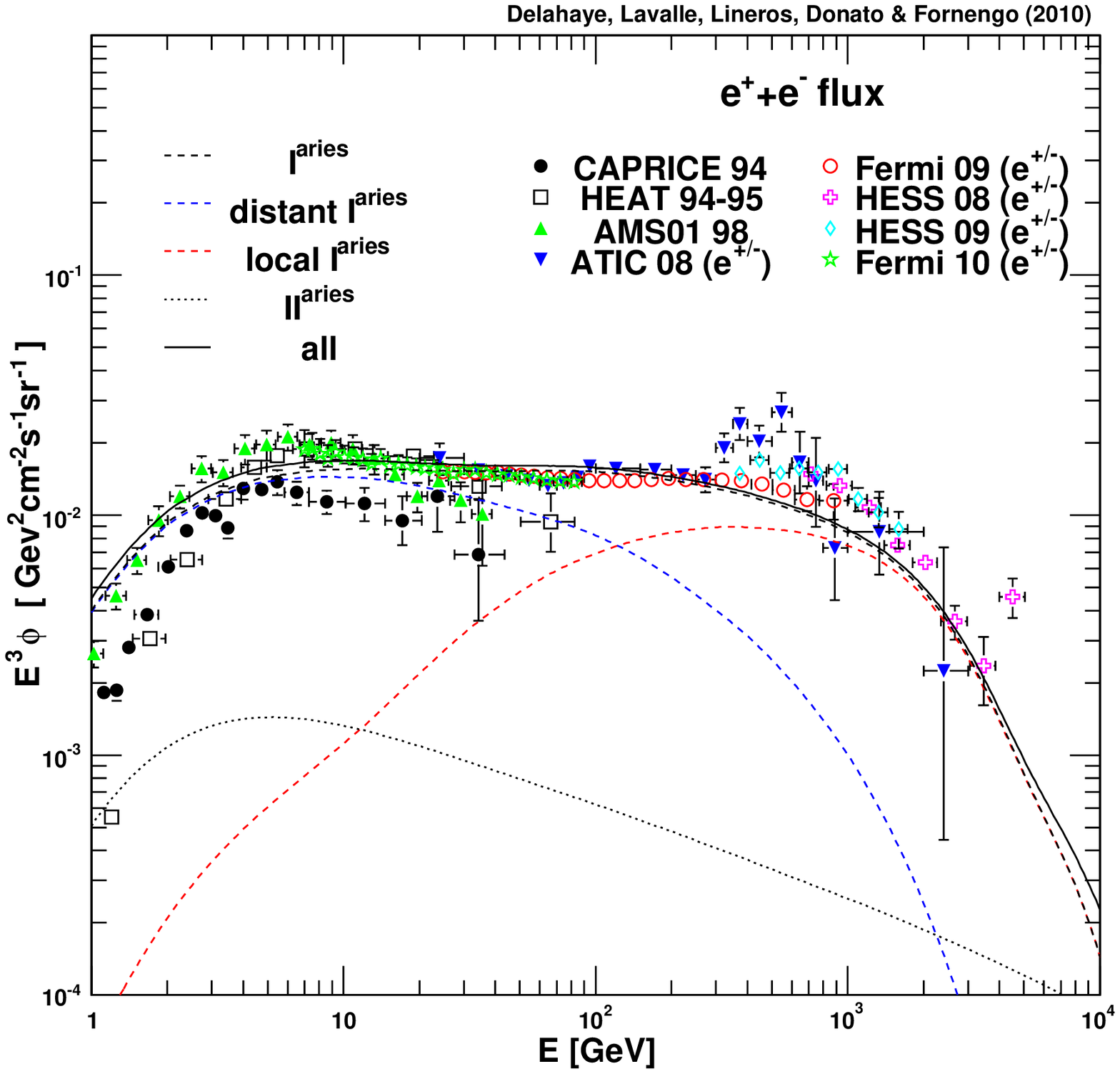}\includegraphics{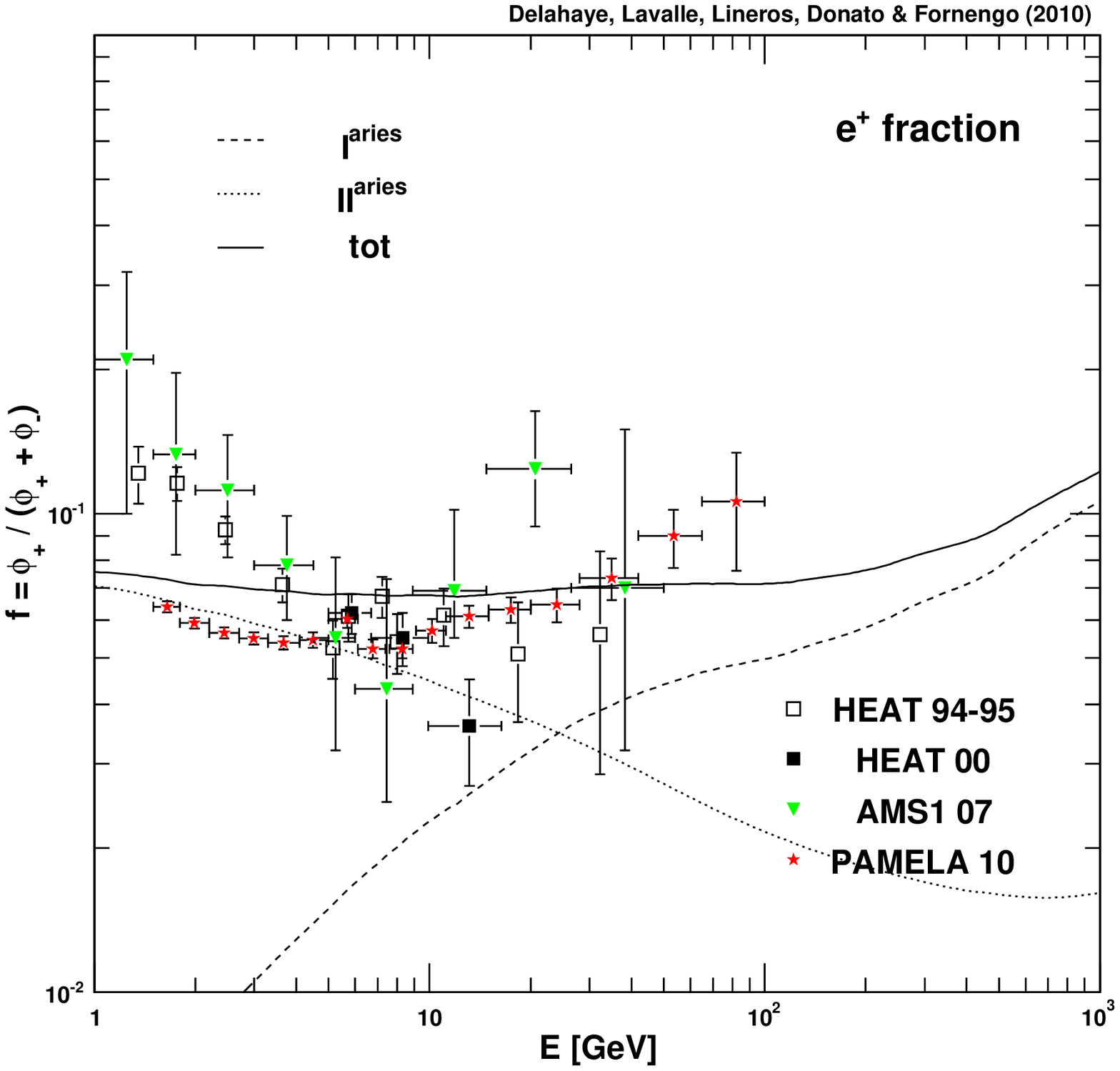}}
\caption{\label{fig3} 
\emph{Electrons} flux (left) and positron fraction (right) versus energy. 
\emph{Electrons} are calculated using information about supernova remnants, pulsars, and also contribution from secondaries.
Both observables are encompassed by the conservative assumption about astrophysical sources.
Further details in~\cite{2010arXiv1002.1910D}.}
\end{fig}
\section{Conclusions}

The cosmic rays anomalies observed by PAMELA and FERMI, have triggered a revolution in our understanding about the galactic environment.
Dark matter is far the most exiting solution to the puzzle, but this supposition considers that the astrophysical background is absolutely known.
Uncertainties in the propagation model and in the sources make harder the task to discriminate a possible non-astrophysical source. 
After some detailed study regarding galactic supernova population, it seems natural that supernova remnants and pulsars may be the solution to these anomalies.
It is indispensable to look for dark matter in other ways, but also to refine the propagation model using other types of observables like diffuse gamma emission and radio observations.

\section*{Acknowledgements}
R.L. is grateful to T. Delahaye, J. Lavalle, N. Fornengo, P. Salati, and F. Donato for the rich collaboration and discussion at different stages of the publications referred in this proceeding. Also, R.L. acknowledges financial support given by Ministero dell'Istruzione, dell'Universit\`a e della Ricerca (MIUR), by the University of Torino (UniTO), by the Istituto Nazionale di Fisica Nucleare (INFN) within the Astroparticle Physics Project, and by the Italian Space Agency (ASI) under contract Nro: I/088/06/0. 

\section*{References}
\bibliographystyle{iopart-num}
\bibliography{bib/refs} 

\end{document}